\PassOptionsToPackage{dvipsnames}{xcolor}
\documentclass[
  twocolumn,
 showpacs,
 showkeys,
 preprintnumbers,
 amsmath,amssymb,
 aps,
 pra,
  longbibliography,
 floatfix,
 ]{revtex4-2}

\usepackage{mathptmx}

\usepackage{amssymb,amsthm,amsmath}

\usepackage{tikz}
\usetikzlibrary{calc,shapes.geometric}
\usepackage {pgfplots}
\pgfplotsset {compat=1.8}
\usepackage{graphicx}

\usepackage{xcolor}

\usepackage{hyperref}
\hypersetup{
    colorlinks,
    linkcolor={blue},
    citecolor={red!75!black},
    urlcolor={blue}
}

\usepackage{mathbbol}

\usepackage{multirow}

\newcommand\myotimes{ }

\begin{document}

\title{Converting nonlocality into contextuality}

\author{Karl Svozil}
\email{karl.svozil@tuwien.ac.at}
\homepage{http://tph.tuwien.ac.at/~svozil}

\affiliation{Institute for Theoretical Physics,
TU Wien,
Wiedner Hauptstrasse 8-10/136,
1040 Vienna,  Austria}

\date{\today}

\begin{abstract}
Matrix pencils provide a robust method for finding simultaneous eigensystems of mutually commuting degenerate operators. In this paper, we utilize these techniques to investigate the quantum logical structures of the Peres-Mermin square and the Greenberger-Horne-Zeilinger-Mermin configuration. Our analysis uncovers analogous complete contradictions between classical and quantum predictions in a four-dimensional system involving two spin-1/2 particles.
\end{abstract}


\maketitle

\section{Introduction}


Heuristically, quantum contextuality encompasses any aspect that contradicts classical predictions, with strong types of contextuality entailing complete contradictions relative to classical expectations. In what follows, we shall concentrate on `strong' quantum contextuality rendered by operator-valued arguments exhibiting nonlocality. While the inverse problem---converting contextuality into nonlocality~\cite{cabello2020converting}---can be of empirical importance, solving the former task can identify the particular type of contextuality exhibited, as well as suggest further experiments.

From a structural standpoint---in terms of the quantum logical algebraic relations of the associated propositions---operator-valued arguments may be closely related, although they may formally appear to be very different. For instance, as observed by Cabello~\cite{cabello-96,Cabello-2013-Hardylike}, Hardy's theorem~\cite{Hardy-92,Hardy-93} can, in quantum logical terms, be transcribed as a true-implies-false arrangement (in graph theoretical terms, a gadget) of observables~\cite{2018-minimalYIYS,svozil-2020-hardy}.
However, as we shall see in comparing Kochen-Specker (KS) and Greenberger-Horne-Zeilinger arguments, there need not be such a close relationship.

Operators in quantum mechanics can have varied spectra, but through their spectral decomposition,
their fundamental components are orthogonal projection operators formed from orthonormal bases of Hilbert spaces.
In this way, every normal operator `masks' or `contains' within itself an orthonormal basis, which
can be identified with a measurement context. From dimension three onwards, these bases can intertwine.

Due to their idempotent nature, projection operators admit only two real solutions (0 and 1) and can thus be associated with elementary logical propositions. The Hilbert space operations of the linear span and the intersection can then be identified with the logical disjunction `or' and conjunction `and' operations, respectively~\cite{birkhoff-36,vonNeumann2018Feb}.

Hence, general quantum observables, which are not elementary binary propositions,
mask the underlying contexts, as they do not directly reflect propositions but rather functions thereof~\cite[\S~82]{halmos-vs}.
To analyze and utilize these observables and their associated operators effectively,
it is necessary to delineate the undelying contexts by extracting their associated idempotent self-adjoint projection operators.
In von Neumann's words~\cite[pp.~241, 242]{vonNeumann2001},
``from the mathematical point of view the more desirable system to treat is not operator theory,
but that part of it which deals with idempotents, because that corresponds to logics,
whereas the whole system corresponds to a somewhat unpleasant extension
of logics, namely where you deal with quantities which can have any number of numerical values,
in other words, physical quantities.''

If the operator is nondegenerate and thus of maximal resolution, this extraction of projection operators is straightforward: It requires computing the eigensystem.



However, if the operators are degenerate, they have multiplicities in the eigenvalues,
which makes their spectral decomposition---and thus their contexts---somewhat arbitrary.
This arbitrariness of the contexts of quantum observables can be overcome
by enlisting other observables that mutually commute with each other and with the original observable.
In this way, a proper `complete' or `maximal resolution' system of observables could uniquely define a context.

A technical problem arises if the mutually commuting operators of the observables are all degenerate.
For the sake of an example take, for instance, the two Hermitian matrices
\[
\begin{pmatrix}
0 &  1 &  0 &  0 \\  1 &  0 &  0 &  0 \\  0 &  0 &  0 &  -1 \\  0 &  0 &  -1 &  0
\end{pmatrix}
\text{ and }
\begin{pmatrix}
0 &  0 &  0 &  -1 \\  0 &  0 &  1 &  0 \\  0 &  1 &  0 &  0 \\  -1 &  0 &  0 &  0
\end{pmatrix}^\intercal,
\]
which commute, and yet, none of their respective eigenvalues coincide: Indeed, the eigensystem of the first matrix consist of separable vectors
$
\begin{pmatrix}
1 ,  \pm 1 ,  0 , 0
\end{pmatrix}^\intercal
$
and
$
\begin{pmatrix}
 0 , 0  , 1 ,  \pm 1
\end{pmatrix}^\intercal
$
while
the eigenvectors of the second matrix
$
\begin{pmatrix}
1  ,  0 , 0 ,  \pm 1
\end{pmatrix}^\intercal
$
and
$
\begin{pmatrix}
 0  , 1 ,  \pm 1 , 0
\end{pmatrix}^\intercal
$
(the symbol $\intercal$ stands for transposition)
 are all nonseparable.
In such cases, finding their respective unique context can be rather tedious, although constructively feasible,
as it involves finding simultaneous eigenvectors for all the commuting operators~\cite[Section 1.3]{Horn-Johnson-MatrixAnalysis}.


In the first part of this paper, we propose applying generalized matrix pencils to the problem of finding a maximal resolution operator for a set of mutually commuting degenerate operators. With this approach, determining the associated unique context becomes straightforward: it only requires computing the eigensystem of a nondegenerate and thus maximal resolution operator that has no eigenvalue multiplicities.

Next, we apply the matrix pencil method to two interesting cases: the Peres-Mermin (PM) square and Mermin's form of the Greenberger-Horne-Zeilinger (GHZM) argument. We find that the quantum logical structure of the PM square constitutes a KS argument, as it does not allow a classical counterfactual truth assignment that is both noncontextual aas well as allows global value assignments of (potential) elements of physical reality. The GHZM argument, on the other hand, is fundamentally different: it operates within a single three-partite context that permits as many classical truth assignments (eight) as its elementary propositions. However, through a parity argument, the associated eigensystem contradicts these classical predictions. This is because a serial concatenation of mutually commuting operators includes only even factors of counterfactual observables, yet their product yields a negative value quantum mechanically.

In the third part, we use the matrix pencil method to construct a two-partite GHZM-type argument. This configuration is based on the observables in the PM square that have a nonlocal eigensystem corresponding to nonseparable vectors associated with entangled quantum states.

\section{Matrix pencils}

The  algorithmic  and thus constructive
analysis of a transcription process for cases of operator-value arguments demonstrating nonclassical behavior is based on the proper spectral decomposition of the operators involved.
Mutually commuting normal operators (such as Hermitian or unitary operators that commute with their respective adjoints)
 $A_1, \ldots, A_l$ share common projection operators.
However, if their spectra are degenerate we need to find an orthonormal basis in which every single one of this collection of mutually commuting operators is diagonal.
The conventional approach to this task can be quite complex~\cite{Nordgren2020Jun}.
Alternatively, we can diagonalize the generalized matrix pencil that is a linear combination of the
operator matrices~\cite[Chapter~12]{Gantmacher2}:
\begin{equation}
P = \sum_{i=1}^{l} a_i A_i,
\label{2024-convert-matrixpencil}
\end{equation}
where $a_i$ are scalars (for our purposes, real numbers).
As $P$ commutes with $A_1, \ldots, A_l$, they share a common set of projection operators.
Moreover, since the scalar parameters $a_i$ can be adjusted, and in particular, can be identified with Kronecker delta functions $\delta_{ij}$,
and as $P$ commutes with each operator $A_j$ for $1 \le j \le l$, $P$ and $A_j$ share a common set of projection operators.

Equipped with these techniques, any collection of commeasurable multipartite observables corresponding
to mutually commuting operators can be transcribed into
projection operators
in the spectrum of the operators of these observables.
If these operators render a maximal resolution, the respective vectors correspond to an orthonormal basis called a context with respect to $A_1, \ldots, A_l$.
The merging or pasting of possibly intertwining contexts then generates a quantum logic which can be
analyzed to identify and characterize the contextual (nonclassical) predictions and features.


\begin{table*}[ht]
\caption{\label{2024-convert-matrixpencil-peres} Eigensystems of the matrix pencils of the rows and columns of the PM square~\eqref{2024-convert-PeresSquare}
with normalization factors omitted.
The eigenvectors corresponding to the last row and column are nonseparable and thus entangled, while all others are separable.
This set of 24 vectors includes the 18 vectors of Cabello, Estebaranz and Garc{\'{i}}a-Alcaine~\cite{cabello-96}.
As already noted by Peres~\cite{peres-91},
these six `primary' contexts associated with orthogonal tetrads are disjoint (not intertwined).
In the hypergraph representation depicted in Figure~\ref{2024-convert-f-24-24}(a) they are represented as the `small ovals' on the six edges of the hypergraph.
}
\begin{ruledtabular}
\begin{tabular}{ccccccc}
matrix pencils&\multicolumn{4}{c}{eigenvalues}\\
&$a - b - c$& $-a + b - c$& $-a - b + c$&   $a + b + c$\\
\hline
$
a \sigma_z  \myotimes   \mathbb{1}_2 + b \mathbb{1}_2 \myotimes   \sigma_z + c \sigma_z  \myotimes   \sigma_z
$ & $
\vert 7 \rangle = \begin{pmatrix}  0, 1, 0, 0\end{pmatrix}^\intercal $ & $
\vert 3 \rangle = \begin{pmatrix}  0, 0, 1, 0\end{pmatrix}^\intercal $ & $
\vert 1 \rangle = \begin{pmatrix}  0, 0, 0, 1\end{pmatrix}^\intercal $ & $
\vert 17 \rangle = \begin{pmatrix}  1, 0, 0, 0\end{pmatrix}^\intercal
$ \\ $
a \mathbb{1}_2 \myotimes   \sigma_x + b \sigma_x  \myotimes   \mathbb{1}_2 + c \sigma_x  \myotimes   \sigma_x
$ & $
\vert 20 \rangle = \begin{pmatrix}  -1, -1, 1, 1\end{pmatrix}^\intercal $ & $
\vert 13 \rangle = \begin{pmatrix}  -1, 1, -1, 1\end{pmatrix}^\intercal $ & $
\vert 11 \rangle = \begin{pmatrix}  1, -1, -1, 1\end{pmatrix}^\intercal $ & $
\vert 24 \rangle = \begin{pmatrix}  1, 1,  1, 1\end{pmatrix}^\intercal
$ \\ $
a \sigma_z  \myotimes   \sigma_x + b \sigma_x  \myotimes   \sigma_z + c \sigma_y  \myotimes   \sigma_y
$ & $
\vert 21 \rangle = \begin{pmatrix}  1, 1, -1, 1\end{pmatrix}^\intercal $ & $
\vert 14 \rangle = \begin{pmatrix}  1, -1, 1, 1\end{pmatrix}^\intercal $ & $
\vert 23 \rangle = \begin{pmatrix}  -1, 1, 1,  1\end{pmatrix}^\intercal $ & $
\vert 10 \rangle = \begin{pmatrix}  -1, -1, -1, 1\end{pmatrix}^\intercal
$ \\ $
a \sigma_z  \myotimes   \mathbb{1}_2 + b \mathbb{1}_2 \myotimes   \sigma_x + c \sigma_z  \myotimes   \sigma_x
$ & $
\vert 12 \rangle = \begin{pmatrix}  -1, 1, 0, 0\end{pmatrix}^\intercal $ & $
\vert 4 \rangle = \begin{pmatrix}  0, 0, 1, 1\end{pmatrix}^\intercal $ & $
\vert 2 \rangle = \begin{pmatrix}  0, 0, -1, 1\end{pmatrix}^\intercal $ & $
\vert 22 \rangle =  \begin{pmatrix}  1, 1, 0, 0\end{pmatrix}^\intercal
$ \\ $
a \mathbb{1}_2 \myotimes   \sigma_z + b \sigma_x  \myotimes   \mathbb{1}_2 + c \sigma_x  \myotimes   \sigma_z
$ & $
\vert 15 \rangle = \begin{pmatrix}  -1, 0, 1, 0\end{pmatrix}^\intercal $ & $
\vert 8 \rangle = \begin{pmatrix}  0, 1, 0, 1\end{pmatrix}^\intercal $ & $
\vert 6 \rangle = \begin{pmatrix}  0, -1, 0, 1\end{pmatrix}^\intercal $ & $
\vert 19 \rangle = \begin{pmatrix}  1, 0, 1, 0\end{pmatrix}^\intercal
$ \\
\hline
&$-a - b - c$& $a + b - c$& $a - b + c$&   $-a + b + c$\\
\hline
$
a \sigma_z  \myotimes   \sigma_z + b \sigma_x  \myotimes   \sigma_x + c \sigma_y  \myotimes   \sigma_y
$ & $
\vert 5 \rangle = \vert \Psi_- \rangle = \begin{pmatrix}  0, 1, -1, 0\end{pmatrix}^\intercal $ & $
\vert 18 \rangle = \vert \Phi_+ \rangle = \begin{pmatrix}  1, 0, 0, 1\end{pmatrix}^\intercal $ & $
\vert 16 \rangle = \vert \Phi_- \rangle = \begin{pmatrix}  1, 0, 0, -1\end{pmatrix}^\intercal $ & $
\vert 9 \rangle = \vert \Psi_+ \rangle = \begin{pmatrix}  0, 1, 1, 0\end{pmatrix}^\intercal
$
\end{tabular}
\end{ruledtabular}
\end{table*}

\section{Peres-Mermin square}

Applying these techniques to the Peres-Mermin (PM) square~\cite{peres111,mermin90b,peres-91,cabello2021contextuality} renders
24 propositions and 24 contexts, henceforth called the 24-24 configuration,
that is the `completion' of the (minimal in four dimensions~\cite{Pavicic-2005,pavicic-2005csvcorri}) 18-9 KS configuration comprising 18 vectors in 9 contexts~\cite{cabello-96}.
In more detail, this configuration involves nine dichotomic observables with eigenvalues $\pm 1$ arranged in a $3 \times 3$ PM matrix~\eqref{2024-convert-PeresSquare}.
Its rows and columns are masking six four-element contexts, one per row and column ($\sigma_i \myotimes \sigma_j$ stands for the tensor product of Pauli spin matrices $\sigma_i \otimes \sigma_j$, with similar notation for $\mathbb{1}_2$)
\begin{equation}
\begin{pmatrix}
\sigma_z \myotimes  \mathbb{1}_2 & \mathbb{1}_2 \myotimes  \sigma_z & \sigma_z \myotimes  \sigma_z \\
\mathbb{1}_2 \myotimes  \sigma_x & \sigma_x \myotimes  \mathbb{1}_2 & \sigma_x \myotimes  \sigma_x \\
\sigma_z \myotimes  \sigma_x & \sigma_x \myotimes  \sigma_z & \sigma_y \myotimes  \sigma_y
\end{pmatrix}
.
\label{2024-convert-PeresSquare}
\end{equation}

To explicitly demonstrate the difficulties involved co-diagonalization of commuting degenerate matrices
consider the last row of  the PM square~(\ref{2024-convert-PeresSquare}).
Its operators
$\sigma_z  \myotimes   \sigma_x$, $\sigma_x  \myotimes   \sigma_z$, and $\sigma_y  \myotimes   \sigma_y $ mutually
commute---for instance, $[\sigma_z  \myotimes   \sigma_x,\sigma_y  \myotimes   \sigma_y]=0$.
However, a straightforward calculation of the eigenvectors of $\sigma_z \myotimes  \sigma_x$ yields:
$\begin{pmatrix}
{0, 1, 0, 1}
\end{pmatrix}^\intercal $,
$\begin{pmatrix}
{-1, 0, 1, 0}
\end{pmatrix}^\intercal $,
$\begin{pmatrix}
{0, -1, 0, 1}
\end{pmatrix}^\intercal $, and
$\begin{pmatrix}
{1, 0, 1, 0}
\end{pmatrix}^\intercal $.
None of these eigenvectors are eigenvectors of $\sigma_y \myotimes  \sigma_y$, and vice versa.
This demonstrates the difficulties involved in co-diagonalizing  commuting degenerate matrices.

Nonetheless, the `joint' PM square contexts are revealed as the normalized eigenvectors of the respective matrix pencils~\eqref{2024-convert-matrixpencil}.
Table~\ref{2024-convert-matrixpencil-peres} enumerates those contexts,
provided that the $\sigma$-matrices are encoded in the standard form.

\begin{figure*}[ht]
\centering
\begin{tabular}{ccc}
\begin{minipage}{.43\textwidth}
\resizebox{1\textwidth}{!}{
\includegraphics{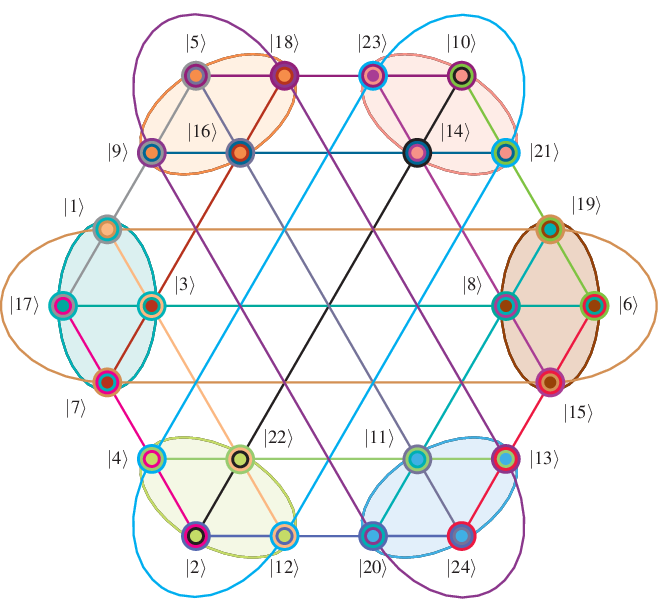}
}
    \end{minipage}%
&
    \begin{minipage}{0.27\textwidth}
\resizebox{1\textwidth}{!}{
\includegraphics{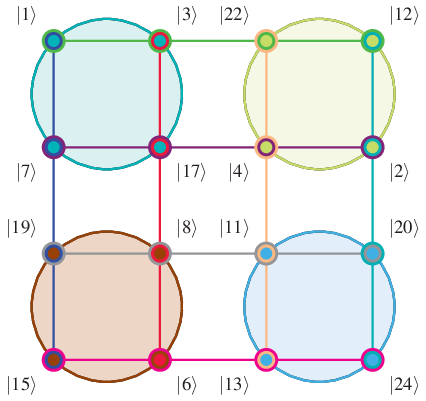}
}
    \end{minipage}%
&
    \begin{minipage}{0.27\textwidth}
\begin{tabular}{c}
\resizebox{1\textwidth}{!}{
\includegraphics{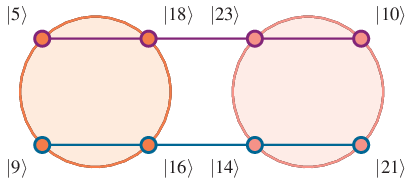}
}
\\
\resizebox{1\textwidth}{!}{
\includegraphics{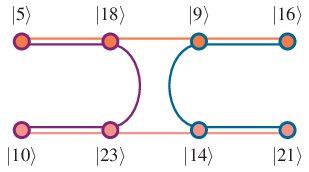}
}
    \end{tabular}
    \end{minipage}%
\\
(a)&(b)&(c)
    \end{tabular}
\caption{\label{2024-convert-f-24-24}
(a)
Hypergraph representing contexts (or cliques or orthonormal bases or maximal operators) 
as unbroken smooth lines. This is an `orthogonal completion'~\cite{peres-91,pavicic-2004ksafq} of the
KS set comprising 18 vectors in 9 contexts introduced by Cabello, Estebaranz and Garc{\'{i}}a-Alcaine~\cite{cabello-96}.
The filled shaded small ovals on the edges correspond to the `primary' isolated (nonintertwined) contexts from the matrix pencil calculations enumerated
in Table~\ref{2024-convert-matrixpencil-peres}.
(b)
Hypergraph representing a 16-12 configuration: 16 elements in 12 contexts enumerated in the first, second, fourth, and fifth row of Table~\ref{2024-convert-matrixpencil-peres}.
These vectors are separable and thus correspond to factorizable, nonentangled states.
(c)
Two equivalent hypergraph representations of a 8-4 configuration---8 elements in 4 contexts
enumerated in the third and sixth row of
Table~\ref{2024-convert-matrixpencil-peres}.
These vectors are nonseparable and thus correspond to entangled states.}
\end{figure*}

Analysis of their orthogonality relations yields an adjacency matrix that, in turn,
can be used to construct the respective  (hyper)graph through the intertwining 24 cliques and thus contexts thereof.
As can be expected, there are only four-cliques corresponding to orthonormal bases in four dimensional Hilbert space.
Figure~\ref{2024-convert-f-24-24}(a) depicts the hypergraph representing these intertwining contexts as unbroken smooth lines,
 and the vector labels as elements of these contexts,
as enumerated in Table~\ref{2024-convert-matrixpencil-peres}.

The 24 rays were already discussed by Peres~\cite{peres-91} as permutations of the vector components of
$\begin{pmatrix}
{1, 0, 0, 0}
\end{pmatrix}^\intercal $,
$\begin{pmatrix}
{1, 1, 0, 0}
\end{pmatrix}^\intercal $,
$\begin{pmatrix}
{1, -1, 0, 0}
\end{pmatrix}^\intercal $,
$\begin{pmatrix}
{1, 1, 1, 1}
\end{pmatrix}^\intercal $,
$\begin{pmatrix}
{1, 1, 1, -1}
\end{pmatrix}^\intercal $, and
$\begin{pmatrix}
{1, 1, -1, -1}
\end{pmatrix}^\intercal $.
The `full' 24-24 configuration was obtained by Pavi\v{c}i\'{c}~\cite{pavicic-2004ksafq} who
reconstructed additional 18 contexts not provided in the original Peres paper~\cite{peres-91} by hand~\cite{pavicic-private-commun-2024}.
Peres' 24-24 configuration is arranged in four-element contexts  associated with four-dimensional Hilbert space, with vector
components drawn from the set $\{-1,0,1\}$, that do not support any two-valued state.
Pavi\v{c}i\'{c}, Megill and Merlet~\cite[Table~1]{pavicic-2010nkss}
have demonstrated that Peres' 24-24 configuration contains 1,233 sets that do not support any two-valued states.
Among these 1,233 sets are six `irreducible' or `critical' configurations which do not contain any proper subset that does not support two-valued states.
Notably, among these configurations is the previously mentioned 18-9 configuration proposed by Cabello, Estebaranz and Garc{\'{i}}a-Alcaine~\cite{cabello-96}.
Previously, Pavi\v{c}i\'{c}, Merlet, McKay, and Megill~\cite[Section~5(viii)]{Pavicic-2005,pavicic-2005csvcorri} had shown that,
among all sets with 24 rays and vector components from the set $\{-1,0,1\}$, and
24 contexts, only one configuration does not
allow any two valued state---and that one is isomorphic to Peres' `full' (including 18 additional contexts)
24-24 configuration enumerated by Pavi\v{c}i\'{c}~\cite{pavicic-2004ksafq}.
This computation had taken one year on a single CPU of a supercomputer~\cite{pavicic-private-commun-2024}.
More recently, Pavi\v{c}i\'{c} and Megill~\cite[Table~1]{Pavii2018} have demonstrated that the vector components from the set $\{-1,0,1\}$
vector-generate a  24-24 set, which contains all smaller KS sets  and is simultaneously isomorphic to the `completed'  24-24 configuration configuration.

We conjecture that if a `larger' collection of contexts (such as 24-24) contains a `smaller' collection of contexts (such as 18-9),
then it inherits the scarcity or total absence of two-valued states of the latter: if the `smaller' set cannot support features related to two-valued states, such as separability of propositions~\cite[Theorem 0]{kochen1},
then intertwining or adding contexts can only impose further constraints, thereby exacerbating the situation by introducing new conditions.

\section{Greenberger-Horne-Zeilinger argument}

Based on the GHZ argument Mermin has suggested~\cite{mermin,mermin90b} a `simple unified form for the major no-hidden-variables theorems' in which he identified four commuting three-partite operators:
$\sigma_x \myotimes  \sigma_x \myotimes  \sigma_x$, $\sigma_x \myotimes  \sigma_y \myotimes  \sigma_y$, $\sigma_y \myotimes  \sigma_x \myotimes  \sigma_y$, and $\sigma_y \myotimes  \sigma_y \myotimes  \sigma_x$.
A parity argument reveals a state-independent quantum contradiction to the classical existence of (local, noncontextual) elements of physical reality:
The quantum mechanical expectation of the product of these four commuting three-partite operators for any quantum state is
$
-1= \langle
-\mathbb{1}_8
 \rangle
=
\langle
\mathbb{1}_2
\myotimes
(
-\mathbb{1}_2
)
\myotimes
\mathbb{1}_2
 \rangle
=
\langle
(
\sigma_x  \cdot \sigma_x  \cdot \sigma_y   \cdot \sigma_y
)
\myotimes
(
\sigma_x   \cdot \sigma_y   \cdot \sigma_x   \cdot  \sigma_y
)
\myotimes
(
\sigma_x   \cdot \sigma_y    \cdot \sigma_y    \cdot \sigma_x
) \rangle
=
\langle  (\sigma_x \myotimes  \sigma_x \myotimes  \sigma_x) \cdot (\sigma_x \myotimes  \sigma_y \myotimes  \sigma_y) \cdot (\sigma_y \myotimes  \sigma_x \myotimes  \sigma_y) \cdot (\sigma_y \myotimes  \sigma_y \myotimes  \sigma_x) \rangle
=
\langle  \sigma_x \myotimes  \sigma_x \myotimes  \sigma_x  \rangle \langle \sigma_x \myotimes  \sigma_y \myotimes  \sigma_y  \rangle \langle
\sigma_y \myotimes  \sigma_x \myotimes  \sigma_y  \rangle \langle   \sigma_y \myotimes  \sigma_y \myotimes  \sigma_x \rangle
$.
In this formulation, every operator $\sigma_x$ and $\sigma_y$ for each of the three particles occurs twice. Therefore, if classically all such single-particle observables would coexist as elements of physical reality and independent of what other measurements are made alongside,
then their respective product must be $1$, the exact negative of the quantum expectation.

Mermin's configuration can be analyzed in terms of its matrix pencil
$
a \sigma_x \myotimes  \sigma_x \myotimes  \sigma_x + b \sigma_x \myotimes  \sigma_y \myotimes  \sigma_y + c \sigma_y \myotimes  \sigma_x \myotimes  \sigma_y + d \sigma_y \myotimes  \sigma_y \myotimes  \sigma_x
$,
thereby revealing the underlying, hidden context in terms of
the simultaneous eigensystem of the four mutually commuting operators.
These eight nonseparable vectors form an orthonormal basis of an eight-dimensional Hilbert space corresponding to an isolated single context~\cite[Table~1]{svozil-2020-ghz} of  entangled states.
Therefore, Mermin's configuration does not constitute a KS proof, as it still permits a separating set of eight two-valued states.

In view of this, how does one arrive at a complete GHZ contradiction with classical elements of physical reality, as outlined above?
The criterion employed in an experimental corroboration~\cite{panbdwz} is to select any one of the eigenstates forming the orthonormal basis, such as
$
(1/\sqrt{2})
\big(
\vert z_+z_+z_+ \rangle  + \vert z_-z_-z_- \rangle
\big)
$.
Since this is an eigenstate of all four terms of the matrix pencil, four separate measurements can be performed (possibly temporally separated) yielding the eigenvalues
$+1$ for
$\sigma_x \myotimes  \sigma_x \myotimes  \sigma_x$
as well as $-1$ for the three others. These three factors  $-1$ and one factor $+1$ contribute to their product value $-1$, in total contradiction to the classical expectation $+1$.
Note that similar contradictions arise if the seven other eigenstates of the matrix pencil are considered~\cite[Table~1]{svozil-2020-ghz}.


\section{Bipartite Greenberger-Horne-Zeilinger argument}

Can an equally convincing argument be made involving just two particles?
Natural candidates would be the `nonclassical' elements of the PM square~\eqref{2024-convert-PeresSquare}.
Note that its `masked' or `hidden' contexts, revealed by the matrix pencils, can be partitioned into four `separable' type contexts
depicted in Figure~\ref{2024-convert-f-24-24}(b)
containing only separable vectors---corresponding to the first and second rows and columns---and two `nonclassical' contexts consisting of nonseparable
vectors---corresponding to the last row and column, as depicted in
Figure~\ref{2024-convert-f-24-24}(c).

Concentrating on these two latter contexts consisting of nonseparable vectors, we make the following observations:
Since the observables from the last row and last column (with the exception of  $\sigma_y \myotimes \sigma_y$) do not commute, they cannot be simultaneously measured.
Nevertheless, by forming products within the last row and column, we may create two commuting operators
%
%
$
(\sigma_z \myotimes \sigma_x) \cdot (\sigma_x \myotimes \sigma_z) = -(\sigma_x \myotimes \sigma_x) \cdot (\sigma_z \myotimes \sigma_z)
=
(\sigma_z \cdot \sigma_x ) \myotimes (\sigma_x \cdot \sigma_z)
=   \sigma_y  \myotimes \sigma_y
= \text{antidiag}
\begin{pmatrix} -1 , 1 ,1, -1
\end{pmatrix}
$.
Their matrix pencil
\begin{equation}
a(\sigma_z \myotimes \sigma_x) \cdot (\sigma_x \myotimes \sigma_z) + b (\sigma_x \myotimes \sigma_x) \cdot (\sigma_z \myotimes \sigma_z)
\label{2024-convert-mppm}
\end{equation}
has a degenerate spectrum with the Bell basis as eigenvectors---the
same as the eigenvectors of the matrix pencil of the last column of the PM square.
(Alternatively, we could have used
the pencil
\(
a(\sigma_z \myotimes \sigma_x) \cdot (\sigma_x \myotimes \sigma_z) + b \sigma_x \myotimes \sigma_x + c \sigma_z \myotimes \sigma_z
\) to avoid multiplicities.)
It is enumerated in Table~\ref{2024-convert-pm-es}.

Hence, preparing a state in one Bell basis state and measuring (successively or separately)
$
(\sigma_z \myotimes \sigma_x) \cdot (\sigma_x \myotimes \sigma_z)
$,
and either
$
(\sigma_x \myotimes \sigma_x) \cdot (\sigma_z \myotimes \sigma_z)
$
or
$
\sigma_x \myotimes \sigma_x
$
as well as
$
\sigma_z \myotimes \sigma_z
$ separately,
yields
\begin{equation}
\begin{split}  -1 =
\langle
-\mathbb{1}_4
 \rangle
=
\langle
\mathbb{1}_2 \myotimes (-\mathbb{1}_2)
 \rangle
\\
=
 \langle
(\sigma_z  \cdot \sigma_x \cdot \sigma_x \cdot \sigma_z ) \myotimes (\sigma_x \cdot \sigma_z \cdot  \sigma_x \cdot \sigma_z)
 \rangle
\\
=
\langle
(\sigma_z \myotimes \sigma_x) \cdot (\sigma_x \myotimes \sigma_z)   \cdot
(\sigma_x \myotimes \sigma_x) \cdot (\sigma_z \myotimes \sigma_z)
\rangle
\\
=
\langle
(\sigma_z \myotimes \sigma_x) \cdot (\sigma_x \myotimes \sigma_z)\rangle
\langle  (\sigma_x \myotimes \sigma_x) \cdot (\sigma_z \myotimes \sigma_z) \rangle
.
\end{split}
\end{equation}

In contrast, and in analogy to Mermin's version of the GHZ argument, the classical prediction is that the product of these terms always needs to be positive, as every alleged `element of reality', in particular corresponding to $\sigma_x$ and $\sigma_z$, enters an even number of times (indeed, twice per particle).

\begin{table}[t]
\caption{\label{2024-convert-pm-es}Eigensystem of the matrix pencil~\eqref{2024-convert-mppm}
associated with the commuting  products of operators in the last (third) row and the last (third) column of the PM square,
constituting the Bell basis.
Inclusion of  $(\sigma_y \myotimes \sigma_y) \cdot (\sigma_y \myotimes \sigma_y) = \mathbb{1}_4$
does not change the calculation and is therefore omitted.
The values $+1$ and $-1$ represent the (co)measured values of the respective commuting operators.
}
\centering
\begin{ruledtabular}
\begin{tabular}{rccccccccc}
\multicolumn{1}{c}{value} &
\multicolumn{1}{c}{vector} &
$(\sigma_z \myotimes  \sigma_x) \cdot (\sigma_x\myotimes  \sigma_z)$ &
$\sigma_x \myotimes  \sigma_x$ &
$\sigma_z\myotimes  \sigma_z$ &
$(\sigma_x \myotimes  \sigma_x) \cdot  (\sigma_z\myotimes  \sigma_z)$
\\
\hline
   $a - b$ &          $ \vert \Psi_+ \rangle $   &   $+1$  &  $+1$    &  $-1$   &  $-1$      \\
   $a - b$ &          $ \vert \Phi_- \rangle $   &   $+1$  &  $-1$    &  $+1$   &  $-1$      \\
   $-a + b$ &         $ \vert \Psi_- \rangle $   &   $-1$  &  $-1$    &  $-1$   &  $+1$      \\
   $-a + b$ &         $ \vert \Phi_+ \rangle $   &   $-1$  &  $+1$    &  $+1$   &  $+1$      \\
\end{tabular}
\end{ruledtabular}
\end{table}

\section{Conclusions}

The matrix pencil method provides an elegant solution for simultaneously diagonalizing commuting operators with degenerate spectra.
It offers a systematic approach for identifying `contextual' nonclassical performance in quantized systems,
particularly in delineating operator-valued arguments.

The Peres-Mermin (PM) square demonstrates a fundamental contradiction (quantum -1 versus classical +1) compared to classical expectations in a dichotomic operator-valued formulation. By employing matrix pencils, this contradiction can be transcribed into a Kochen-Specker (KS) type argument involving 24 vectors. This configuration, which does not support any binary (two-valued) state, consists of 6 `original' isolated contexts from the matrix pencils associated with every row and column of the PM square, as well as 18 `secondary' intertwining contexts obtained by studying orthogonalities.

Mermin's rendition of the GHZ operator-valued argument is fundamentally different. When transcribed into quantum logic, it reveals a single isolated context that is perfectly set-representable, for instance, by partition logic. The quantum state becomes crucial for any experimental corroboration: if one takes any eigenstate of the matrix pencil, it leads to a complete contradiction (again quantum -1 versus classical +1) when multiplying all the results and comparing the squares of operators in a parity argument.

In analyzing the `entangled contexts' corresponding to the last row and column of the PM square and constructing
mutually commuting products thereof, one arrives at a similar argument to Mermin's rendition of the GHZ argument.
It is also state-independent and operates within a single context. The operators involved are:
$
(\sigma_z \myotimes \sigma_x) \cdot (\sigma_x \myotimes \sigma_z)
$  and alternatively, either
$
(\sigma_x\myotimes \sigma_x) \cdot (\sigma_z\myotimes \sigma_z)
$
or
$
\sigma_x\myotimes \sigma_x
$
and
$\sigma_z\myotimes \sigma_z
$
and, although not needed for the constraction,
$(\sigma_y \myotimes \sigma_y) \cdot (\sigma_y\myotimes \sigma_y)$.
These operators commute, and for the Bell basis yield a complete contradiction (quantum -1 versus classical +1)
contingent on the assumption of noncontextual classical existence of those elements of physical reality.
This reduces the eight-dimensional argument to a four-dimensional one.

Why or how can operator-valued contradictions arise within a single isolated context?
This occurs because measurements such as \(\sigma_x \myotimes \sigma_z \),
which partly define a context derived from a matrix pencil,
should not be considered `local' and cannot be conducted as independent single-qubit local measurements~\cite{cabello2021contextuality}.
Such operator-valued arguments are traditionally rooted in the classical assumption that any multi-particle state
can be decomposed into single-particle states while preserving the properties of the original multi-particle state.
However, this assumption fails in the case of entangled states,
which encodes relational information at the expense of local properties~\cite{zeil-99}.
From this perspective, both dichotomic operator-valued GHZM
arguments and binary two-valued state KS arguments against noncontextuality share a nonoperational and therefore
(meta)physical presumption: the contingent use of counterfactuals.

In summary, this paper proposed using matrix pencils to find a maximal resolution operator for mutually commuting degenerate operators, simplifying the determination of the associated unique context. This method was applied to the Peres-Mermin (PM) square and Mermin's form of the Greenberger-Horne-Zeilinger (GHZM) argument. The PM square's quantum logical structure presented a Kochen-Specker (KS) argument, while the GHZM argument involved a single three-partite context, with quantum mechanical results contradicting classical predictions through a parity argument. Finally, the matrix pencil method was used to construct a two-partite GHZM-type argument based on nonlocal observables in the PM square.

\begin{acknowledgments}

This research was funded in whole or in part by the Austrian Science Fund (FWF), Grant-DOI: 10.55776/I4579. For open access purposes, the author has applied a CC BY public copyright license to any author accepted manuscript version arising from this submission.

A question by Bruno Mittnik stimulated this research.
Philippe Grangier drew my attention to von Neumann's address at the International Congress of Mathematicians in Amsterdam, delivered in September 1954~\cite{vonNeumann2001}.
I acknowledge explanations from Alastair Abbott, Costantino Budroni, Ad\'an Cabello and Jan-\AA{}ke Larsson regarding aspects of Reference~\cite{cabello2021contextuality},
as well as explanations, discussions, and suggestions from Mladen Pavicic regarding the properties of the 24-24 configuration.
(Any remaining confusion remains solely with the author.)
\end{acknowledgments}

\bibliography{svozil}

\end{document}